\documentclass[%
reprint,
unsortedaddress,
nofootinbib,
amsmath,amssymb,
aps,
pra,
]{revtex4-2}

\usepackage{graphicx}
\usepackage{amssymb,amsmath}
\usepackage[utf8]{inputenc}
\usepackage{multirow}
\usepackage{xcolor}
\usepackage{hyperref}
\hypersetup{colorlinks=true,linkcolor=blue,anchorcolor=blue,citecolor=magenta,filecolor=blue,urlcolor=blue,bookmarksnumbered=true}
\usepackage[flushleft]{threeparttable}
\usepackage{mathtools}

\usepackage{dsfont}
\usepackage{amsthm}

\usepackage{enumitem}
\setlist[itemize]{leftmargin=10mm}

\usepackage{multirow}

\usepackage{braket}

\usepackage{mathrsfs}

\usepackage{booktabs}

\usepackage{natbib}

\usepackage[ruled,vlined]{algorithm2e}

\begin{document}

%\preprint{APS/123-QED}

\title{Quantum-classical framework for many-fermion response and structure} 

\author{Weijie Du\textsuperscript{1,2}}

\author{Yangguang Yang\textsuperscript{1}}

\author{Zixin Liu\textsuperscript{1,3}}
\email[Email: ]{zixinliu@gdlhz.ac.cn}

\author{Chao Yang\textsuperscript{4}}

\author{James P. Vary\textsuperscript{2}}

\affiliation{\textsuperscript{1}Institute of Modern Physics, Chinese Academy of Sciences, Lanzhou 730000, China}
\affiliation{\textsuperscript{2}Department of Physics and Astronomy, Iowa State University, Ames, Iowa 50010, USA}
\affiliation{\textsuperscript{3}Advanced Energy Science and Technology, Guangdong Laboratory, Huizhou 516000, China}
\affiliation{\textsuperscript{4}Applied Mathematics and Computational Research Division, Lawrence Berkeley National Laboratory, Berkeley, California 54720, USA}

%\author{Weijie Du, Zixin Liu, Yangguang Yang,{$^{\ast}$} Chao Yang, and James P. Vary}

%\email[Email: ]{duweigy@gmail.com}

\date{\today}

\begin{abstract}

Response functions are key observables for probing the structure and dynamics of many-body systems. 
We introduce and demonstrate a quantum-classical framework for computing response functions of general many-fermion systems that also provides the full bound-state spectrum. 
The framework employs the Lorentz integral transform and a new Hamiltonian input scheme that enables practical and scalable circuit constructions for general many-fermion Hamiltonians.
Within this framework, we develop a hybrid strategy to evaluate the Lorentz integral and propose three protocols to extract response functions and bound-state structural information. 
As a demonstration, we apply the method to \({}^{19}\mathrm{O}\) with realistic internucleon interactions, computing both the bound-state spectrum and the response function. 
We envision that our approach will open new avenues for exploring the structure and dynamics of a broad class of many-body systems across diverse fields.

\end{abstract}

\maketitle

{\bf Introduction.--}
%{\color{magenta} \it  Importance of the Response and Structure:}
The structure and dynamics of quantum many-body systems are fundamental to understanding strongly interacting systems across quantum chemistry \cite{cook2012handbook,bauer2020quantum}, nuclear physics \cite{RevModPhys.77.427,hjorth2017advanced,Barrett:2013nh,goldberger2004collision,wong1998introductory}, high-energy physics \cite{peskin2018introduction,BRODSKY1998299}, and condensed matter physics \cite{fradkin2013field,michael2005computational}. Among the key observables are response functions \cite{RevModPhys.87.1067,stefanucci2025nonequilibrium,fetter2003quantum}, which encode rich information about both the dynamical and structural properties of these systems. Response functions define cross sections in scattering experiments \cite{PhysRevLett.127.072501,PhysRevLett.111.122502,PhysRevC.90.064619,Quaglioni:2007eg,PhysRevLett.134.202501,PhysRevLett.134.192701,BERTSCH1975125} involving photons, electrons, neutrinos and nucleons. They also reveal essential structural information about the underlying quantum systems.

%{\color{magenta} \it  Difficulties:}
While response functions are important and useful quantities, their calculation is challenging. Such computations require access to both bound and continuum states, which obey different boundary conditions and therefore demand distinct numerical techniques, making a unified and consistent treatment difficult. Moreover, the number of many-body states grows exponentially with both the single-particle (SP) basis size and the particle number. This poses a significant challenge for first-principles evaluations of response functions in many-body systems on classical computers.

%{\color{magenta} \it  Introduction of  existing algorithms, and necessity for developing algorithms for the response function:}
Quantum computers hold the promise for addressing quantum many-body problems that are intractable on classical computers \cite{feynman1982simulating}. Research efforts have been devoted to solving eigenvalue problems, including the variational \cite{Cao_2019,McArdle_2020,Tilly_2022,Yuan_2019}, subspace-expansion-based \cite{Yoshioka_2025,yu2025quantumcentricalgorithmsamplebasedkrylov,Du:2024zvr,motta2023subspace,Kirby_2023,Epperly_2022,PhysRevA.107.032414,PhysRevA.105.022417,Stair_2020}, and imaginary-time-evolution \cite{motta2020determining,Yeter-Aydeniz:2019htv} quantum eigensolvers. Meanwhile, quantum algorithms have also been proposed for simulating real-time evolution in many-body systems, such as the Trotter-based methods \cite{lloyd1996universal,PhysRevLett.79.2586}, truncated Taylor series \cite{PhysRevLett.114.090502}, truncated Dyson series \cite{PhysRevA.99.042314}, qubitization \cite{Low_2019}, time-dependent qubitization \cite{PRXQuantum.5.040316}, qDrift \cite{PhysRevLett.123.070503}, time-dependent qDrift \cite{Berry_2020}, and Magnus-expansion-based method \cite{Fang_2025,fang2025highordermagnusexpansionhamiltonian}. Few approaches \cite{Weiss_2025,lpz2-j7vg,Kokcu:2023vwg,PhysRevD.105.074503,PhysRevA.102.022408,PhysRevA.102.022409,Roggero:2018hrn} have been proposed for quantum computing response functions that relate directly to reaction cross sections of realistic, strongly interacting systems; existing works are mainly based on time-evolution-based methods and have been demonstrated only for toy models, highlighting the need for practical and scalable algorithms applicable to response functions of realistic many-body systems.

%{\color{magenta} \it Our approach:}
In this work, we develop a quantum-classical framework for computing nuclear response functions, which also provides bound-state structural information. Our approach is based on the Lorentz integral transform (LIT) \cite{Efros_1994,Efros_2007}, which has been widely applied to scattering processes such as nuclear photoabsorption and electroweak reactions \cite{PhysRevLett.127.072501,PhysRevLett.111.122502,PhysRevC.90.064619,PhysRevLett.134.202501,Quaglioni:2007eg} on classical computers. Within the LIT framework, the response function is mapped to a Lorentz integral (LI) via convolution with a smooth kernel. The LI can be evaluated by solving inhomogeneous many-body Schr\"odinger equations using only bound-state techniques. The response function is then extracted via the integral inversion.

Although conceptually powerful, the LIT approach is demanding in computational resources as it necessitates treating many-body problems on classical computers. We present a hybrid scheme that efficiently evaluates the LI for many-body systems using the Chebyshev polynomial expansion (CPE) \cite{fox1968chebyshev,boyd2013chebyshev}, where the evaluation of the LI reduces to quantum computation of a limited set of Chebyshev moments (CMs) determined by the many-body Hamiltonian and the system-probe interaction. We further propose protocols for computing the response functions and full bound-state spectra of many-body systems.

Our framework also employs a new Hamiltonian input scheme developed in our previous work \cite{Du:2025lhq}, which enables practical and scalable circuit constructions for general second-quantized many-fermion Hamiltonians. Based on the ideas of Boolean masking and discrete-time quantum walks \cite{PhysRevLett.102.180501,childs2010relationship,berry2012black,Childs_2017}, this efficient input scheme avoids the expensive compilation of second-quantized Hamiltonians for many-fermion systems and the intricate oracle constructions required by standard input methods (e.g., Jordan--Wigner \cite{JordanWigner1928,Nielsen2005TheFC} or Bravyi--Kitaev encodings \cite{Bravyi_2002,Seeley_2012} implemented with LCU \cite{Childs2012HamiltonianSU,PhysRevLett.114.090502,RyanBabbushNJP2016}). By employing quantum signal processing \cite{PhysRevLett.118.010501,gilyen2019quantum}, our input scheme enables efficient evaluation of the CMs and, consequently, the LI.

%{\color{magenta} \it Merits of our algorithm:}
Our hybrid framework establishes a practical and scalable approach for computing response functions of general many-fermion systems using future fault-tolerant quantum hardware. Moreover, it enables access to complete bound-state spectra of many-fermion systems, surpassing the above-mentioned quantum eigensolvers that target only a limited set of eigenenergies (e.g., the ground-state energy via VQE \cite{Tilly_2022}). To the best of our knowledge, approaches that enable combined response-function and structural studies of realistic many-fermion systems on quantum computers remain largely unexplored.

%{\color{magenta} \it Demonstration and perspectives:}
We demonstrate the framework by computing the full bound-state spectrum and response function of \({}^{19}\mathrm{O}\) using a realistic strong-force interaction \cite{Shin:2024zpe} developed from fundamental theories \cite{Epelbaum:2008ga,Epelbaum:2014efa,Epelbaum:2014sza}. 
Our framework is directly applicable to other strongly interacting systems, such as those in hadronic structure and dynamics (e.g., within the BLFQ approach \cite{Vary:2025yqo,Honkanen:2010rc,Vary:2009gt}), when combined with consistent input scheme for {many-boson} Hamiltonians \cite{Du:2024ixj}.

{\bf Theory.--}
%{\color{magenta} \it Many-fermion Hamiltonian.--}
The second-quantized Hamiltonian of a general many-fermion system is
\begin{equation}
	H = \sum _j \langle Q_j | H | P_j \rangle b^{\dagger} _{Q_j} b_{P_j},
	\label{eq:2nd_quantized_Hamiltonian}
\end{equation} 
where we index each monomial by $j \in [0, \mathcal{D}-1]$. We denote $Q_j \mapsto \{ p_j, q_j , \cdots r_j \}$ and $P_j \mapsto \{ u_j, v_j, \cdots , w_j \}$, where we order the SP bases as $p_j <q_j < \cdots < r_j$ and $ u_j < v_j < \cdots < w_j $. 
The {\it few-body matrix element} is $ \langle Q_j | H | P_j \rangle \equiv \langle  p_j, q_j, \cdots r_j | H | u_j, v_j, \cdots w_j \rangle $.
We have $ b^{\dagger}_{Q_j} \equiv a^{\dagger} _{p_j} a^{\dagger} _{q_j} \cdots a^{\dagger} _{r_j} $ and $  b_{P_j} \equiv a_{w_j} \cdots a_{v_j} a_{u_j} $. The anticommutation relations hold for the fermion operators, i.e., $\{ a_p, a^{\dagger}_q \} = \delta _{p,q}$, $\{a_p, a_q \} = \{ a^{\dagger}_p , a^{\dagger}_q \} =0$. The fermion operators act on the occupation modes of SP basis ``$p$" as $a^{\dagger}_p \ket{0}_p = \ket{1}_p$, $a_p \ket{1}_p = \ket{0}_p$,  and $a^{\dagger}_p \ket{1}_p = a_p\ket{0}_p = 0$.

%{\color{magenta} \it The Schr\"odinger equation.--}
The Schr\"odinger equation of the many-body system reads
\begin{equation}
	(H- E_{n}) \ket{\Psi _{n}} = 0 ,
\end{equation}
where $\ket{\Psi _{n}}$ and $E_{n}$ (with $n=0, 1, 2, \cdots $) denote the eigenstate and corresponding eigenenergy, respectively.

The response function $R(E)$ describes the system’s response to an external perturbative probe that transfers energy $E$ to it. $R(E)$ admits the form \cite{Efros_2007}
\begin{equation}
	R(E) =  \ \mathclap{\displaystyle\int}\mathclap{\textstyle\sum} \ \ dn \langle \Psi _0 | V^{\dagger} | \Psi _{n} \rangle \langle \Psi_{n} | V | \Psi _0 \rangle \delta (E_{n} - E) ,
	\label{eq:response_function_formula}
\end{equation}
with $V$ denoting the system-probe interaction operator. Combined with the factor that describes the kinematics, $R(E)$ produces the reaction cross section. For clarity, we refer $\ket{ \Omega } = V \ket{\Psi _0}$ as the ``source state". 

While $R(E)$ is a useful quantity, its first-principles calculations for many-body systems is generally challenging. Two major difficulties arise: (1) $R(E)$ depends on a large number of bound and continuum states $\{\ket{\Psi_n}\}$ subject to different boundary conditions, making their unified and consistent treatment computationally difficult; and (2) first-principles calculations for quantum many-body systems are computationally demanding on classical computers.

%{\color{magenta} \it The LIT method.--}
We address the first difficulty by utilizing the LIT  \cite{Efros_2007} method, which presents a practical approach to consistently treating the bound and continuum states in solving $R(E)$. In particular, one considers an integral transform 
\begin{equation}
	\mathcal{ L} (\sigma) =  \ \mathclap{\displaystyle\int}\mathclap{\textstyle\sum} \ \ K(\sigma ,E) R(E) dE .
	\label{eq:integral_transform}
\end{equation}
We take $K(\sigma , E) $ to be the Lorentzian kernel $ \big[ ( \sigma ^{\ast} -E ) ( \sigma - E ) \big]^{-1}$, where $\sigma = x + \sigma _R + i \sigma _I$ with $\sigma _R \in \mathbb{R}$ and $\sigma _I > 0$. 
In the following discussions, we denote the function $ f(\sigma _R$, $\sigma _I, x)$ as $f(\sigma)$ for simplicity. In contrast to existing literature~\cite{Efros_2007}, we introduce a free parameter $x$ in this work, where $x$ takes only two values, $0$ and $E_0$. The utility of $x$ will become clear in the following discussion. We also note that various kernel choices have been considered in the literature (see, e.g.,~\cite{Efros_2007, PhysRevA.102.022409, PhysRevE.105.055310, Zhang:2024ril, Zhang:2024gac}), leading to different types of integral transformations; our framework discussed below is also adaptable to those cases.

The LI can be rewritten as
\begin{equation}
	\mathcal{ L} (\sigma ) = \ \mathclap{\displaystyle\int}\mathclap{\textstyle\sum} \ \  \frac{ R(E) }{(E- x - \sigma _R)^2 + \sigma _I^2}  dE  =  \langle \widetilde{\Psi}  | \widetilde{\Psi}  \rangle ,
	\label{eq:kernel_solution}
\end{equation}
where $\ket{\widetilde{\Psi} }$ can be solved from the inhomogeneous Schr\"odinger equation
\begin{equation}
	(\sigma - H) \ket{ \widetilde{\Psi} }  =  \ket{\Omega} . 
	\label{eq:inhomo_1} 
\end{equation}
It can be shown that $\ket{\widetilde{\Psi}}$ is localized and has a finite norm~\cite{Efros_2007}. Consequently, both the bound and continuum states of $H$ are encoded within $\ket{\widetilde{\Psi}}$ that can be treated with bound-state-type techniques. This avoids the explicit handling of the distinct asymptotic behaviors of the system's eigenstates.

Different from prototypical approaches \cite{Efros_2007} in solving $\ket{\widetilde{\Psi}}$, we utilize the CPE to evaluate the Green’s function $G(\sigma, H) \equiv (\sigma - H)^{-1}$. This facilitates the introduction of quantum computing to handle the classically demanding many-body calculations in our hybrid approach. In particular, the CPE of $G(\sigma, H)$ is
\begin{equation}
	{G}( \sigma , H) =  \sum _{k=0} f_{ k}(\sigma ) T_k ( H ) ,
	\label{eq:expansion_scalar1}
\end{equation}
where $T_k(\cdot)$ is the Chebyshev polynomial of the first kind \cite{arfken2013mathematical}, with $\|H\|_2 \in [-1,1]$  (one rescales $H$ otherwise) and $k =0, 1, 2, \cdots$. The expansion coefficient $f_{ k }(\sigma )$ is
\begin{equation}
	f_k (\sigma) = \frac{2-\delta_{k,0}}{\sqrt{\sigma ^2 -1}} \big( \sigma -\sqrt{\sigma ^2 -1} \big)^k .
\end{equation}
The summation is truncated according to the required numerical precision. With $\sigma _ I>0$, $G(\sigma, H)$ is smooth for any $\sigma_R \in \mathbb{R}$, as $H$ is Hermitian and its eigenvalues are real. Therefore, the CPE in Eq. \eqref{eq:expansion_scalar1} converges exponentially \cite{10.1007/BF01386223,boyd2013chebyshev,fox1968chebyshev}. 

Compared to our previous work \cite{Du:2025lhq}, where the Green's function is approximated using a discrete Fourier transform combined with the CPE, the polynomial approximation [Eq.~\eqref{eq:expansion_scalar1}] provides a more direct and convenient treatment of the Green's function, in which the singularities are removed by introducing a finite positive $\sigma_I$. The polynomial approximation employed in this work can be further improved by the kernel polynomial method \cite{RevModPhys.78.275} for enhanced numerical precision and stability.

Based on the solution $  \ket{ \widetilde{\Psi} } = {G}( \sigma , H ) \ket{\Omega } $, we have
\begin{multline}
	\mathcal{L} (\sigma )  = \sum _{k=0} \sum _{j=0}  f^{\ast}_{ k }(\sigma )  f_{ j}(\sigma ) \\ 
	\times \frac{1}{2} \big( \langle \Omega | T_{j+k} (H) | \Omega \rangle + \langle \Omega |T_{| j-k |} (H) | \Omega \rangle \big) ,
	\label{eq:result_1_straight}
\end{multline}
where we have employed the identity $ 2 T_j ( \varepsilon )  T_k ( \varepsilon ) = T_{j+k} (\varepsilon) + T_{| j-k |} (\varepsilon) $ with $\varepsilon \in [-1,1]$ \cite{arfken2013mathematical}. In this sense, the LI is expressed as a linear combination of a limited set of CMs $ \{  \bra{\Omega} T_n ( H )  \ket{ \Omega }  \} $. These CMs only depend on $H$ and $V$, which determine the reaction dynamics.

The LI can be separated into two parts
\begin{equation}
	\mathcal{L}(\sigma ) = \underbrace{ \sum _{n} \frac{R_n}{ (\sigma _R - e_n)^2 + \sigma _I^2 } }_{\mathcal{L}_D(\sigma )} + \underbrace{ \int _{e_{\rm th}} ^{\infty } d e \frac{R(e)}{ ( \sigma _R - e )^2 + \sigma ^2_I } }_{ \mathcal{L}_C ( \sigma ) } ,
	\label{eq:response_function_formula_1}
\end{equation}
where $\mathcal{L}_D (\sigma) $ and $\mathcal{L}_C(\sigma )$ denote the contributions from the bound and continuum states, respectively. 
Notably, we define $e_n = E_n - x$, $e = E - x$, and $e_{\rm th} = E_{\rm th} - x$. For $x = E_0$, this notation accounts for excitation energies. Accordingly, we rewrite $R(E)$ as $R(e)$, whereby $\delta(E_n - E)$ is replaced by $\delta(e_n - e)$ and the integration limits are shifted consistently.

\begin{figure}[ht]
	\centering
	\includegraphics[width=0.95\linewidth]{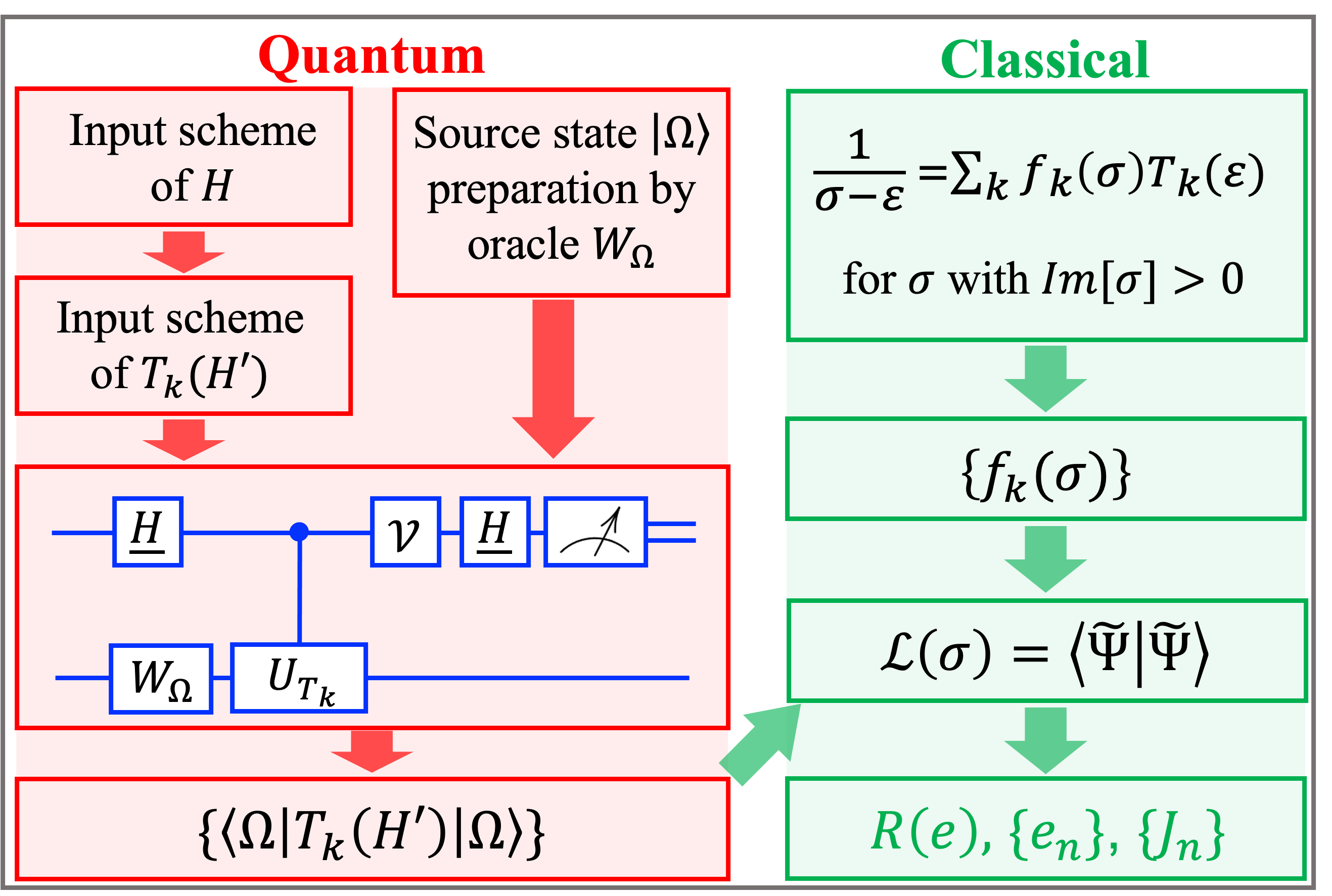}
	\caption{Workflow of the quantum--classical algorithm for solving the response function and bound-state spectrum. The circuit implements a standard Hadamard test \cite{nielsen2010quantum}, with the Hadamard gate denoted by $\underline{H}$. The gate $\mathcal{V}$ is set to the identity for $\mathrm{Re}[\langle \Omega | T_k(H') | \Omega \rangle]$ and to $S^{\dagger}$ for $\mathrm{Im}[\langle \Omega | T_k(H') | \Omega \rangle]$. 
	}
	\label{fig:hybrid_scheme}
\end{figure}

{\bf Quantum-classical framework.--}
To address the second difficulty, namely the exponential growth of computational resources required for first-principles many-body calculations, we propose a hybrid quantum--classical framework, whose workflow is illustrated in Fig.~\ref{fig:hybrid_scheme}. In this framework, quantum computers are employed to evaluate the CMs, which encode essential many-body physics and are computationally demanding on classical computers. These CMs are subsequently processed on classical computers to compute the structure and response functions, where this post-processing can be more efficiently performed classically than on quantum hardware.

On the quantum computers, we employ the \textit{direct encoding scheme} to map $N_{\rm sp}$ SP bases onto $N_{\rm sp}$ qubits, where each qubit records the occupation of the corresponding SP basis. The state of the $q$th qubit is $\ket{b_q}$, with $b_q \in \{0,1\}$, denoting that the $q$th SP basis is vacant ($b_q = 0$) or occupied ($b_q = 1$), respectively. Accordingly, a many-fermion state is encoded as $\ket{\mathcal{F}} = \ket{b_0} \ket{b_1} \cdots \ket{b_{N_{\rm sp}-1}}$, which corresponds to a binary string on the quantum register.

We employ a new Hamiltonian input scheme proposed in our previous work \cite{Du:2025lhq} to encode $H$ [Eq. \eqref{eq:2nd_quantized_Hamiltonian}] on a quantum computer. This scheme constructs circuit representations of the fermionic operators, as well as their various combinations, based on the idea of Boolean masking \cite{helgaker2014molecular}, thereby avoiding expensive compilation overhead associated with the conventional Jordan–Wigner \cite{JordanWigner1928,Nielsen2005TheFC} or Bravyi–Kitaev \cite{Bravyi_2002,Seeley_2012} encodings. With these circuit representations, the quantum-walk approach \cite{PhysRevLett.102.180501,childs2010relationship,berry2012black} is employed to block-encode $H$ as [Eq. (8), \citenum{Du:2025lhq}]
\begin{equation}
	\big(\bra{\mathcal{G}}_{s} \otimes \bra{0} _{a} \big) \mathcal{T}_b^{\dagger} \mathcal{T} _f \big( \ket{\mathcal{F}} _s \otimes \ket{0} _a \big) = \frac{1}{\mathcal{B} \Xi } \bra{\mathcal{G}} H \ket{\mathcal{F}} ,
	\label{eq:block_encoding}
\end{equation} 
where the subscripts $s$ and $a$ denote the system and ancilla registers, respectively. The states $\ket{\mathcal{G}}$ and $\ket{\mathcal{F}}$ encode the many-fermion states in the system register. $\mathcal{T}_b$ and $\mathcal{T}_f$ denote the circuit operations used to construct the quantum-walk states based on $\ket{\mathcal{G}}$, $\ket{\mathcal{F}}$, and $H$. The factor $\mathcal{B} \geq \mathcal{D}$ arises from the diffusion operators in the quantum circuit, whereas $\Xi \geq \max_j \big| \bra{Q_j} H \ket{P_j} \big|$ with $j \in [0, \mathcal{D}-1]$ denotes the maximum norm of the input few-body matrix elements in Eq. \eqref{eq:2nd_quantized_Hamiltonian}. 

By employing a specific technique to encode few-body matrix elements, this quantum-walk-based Hamiltonian input scheme provides a practical circuit design that avoids the explicit construction of the {``Prepare oracle"} required in the standard LCU approach  \cite{Du:2025lhq}, which can become challenging due to the need to extract a large number of matrix elements in general many-body systems. Interested readers are referred to Fig.~2 of Ref.~\cite{Du:2025lhq} for an explicit example of the circuit implementation of the Hamiltonian input scheme.

Based on Eq.~\eqref{eq:block_encoding}, we block-encode the Chebyshev polynomial $T_k(H')$, with $H' = H/(\mathcal{B} \Xi)$, utilizing quantum signal processing \cite{PhysRevLett.118.010501,gilyen2019quantum,lin2022lecture}. For clarity, we denote the unitary that block-encodes $T_k(H')$ as $U_{T_k}$ in Fig.~\ref{fig:hybrid_scheme}. 
A circuit implementation of $U_{T_k}$ can be found, e.g., in Fig.~S1 of Ref.~\cite{Du:2025lhq}.

Our input scheme enables efficient circuit construction for encoding general many-fermion Hamiltonians with low gate cost~\cite{Du:2025lhq}. In particular, the gate complexity scales as $\widetilde{O}(N_{\rm sp}^{2\eta+1})$ for encoding Hamiltonians containing interaction terms up to $\eta$-body operators, where the number of monomials scales as $\widetilde{O}(N_{\rm sp}^{2\eta})$. The number of qubits required to encode a many-fermion Hamiltonian in a basis of $N_{\rm sp}$ SP states is $N_{\rm sp} + \lceil \log_2 \mathcal{D} \rceil + 4$.
Accordingly, $T_k(H')$ is encoded using quantum signal processing with one additional qubit, and the corresponding gate complexity scales as $\widetilde{O}(k \, N_{\rm sp}^{2\eta+1})$.

In this work, we propose an oracle $W_{\Omega}$ to encode the source state $\ket{\Omega} = V \ket{\Psi_0}$ on the system register as $\ket{\Omega }_s = W_{\Omega} \ket{0}_s$. The systematic circuit design of $W_{\Omega}$, consistent with our Hamiltonian input scheme, requires further developments, which we defer to our forthcoming work \cite{ZixinLiu2025unpub}. 

We employ the Hadamard test \cite{nielsen2010quantum} to quantum compute the CMs based on the circuit representations of $T_k(H')$ and $\ket{\Omega}$, which are realized by the input scheme of $T_k(H')$ and the oracle $W_{\Omega}$.

On the classical computers, we collect the CMs evaluated on quantum computers, which are combined with the expansion coefficients to compute the LI [Eq.~\eqref{eq:result_1_straight}]. 
 Based on the LI, we then introduce three protocols to extract the full bound-state spectrum (including the eigenenergies and their total angular momenta $J$) as well as the response function. The details of these oracles are as follows.

{\it The first protocol} determines the structural information of the Hamiltonian. Since the input schemes for both the Hamiltonian and the corresponding Chebyshev polynomials respect the underlying symmetries of the Hamiltonian, the calculations can be carried out within the cascading \(M_J\) scheme~\cite{Barrett:2013nh,Du:2025lhq}. 
In particular, if one elects the source state $\ket{\Omega _{M_J}}$ to be a single-configuration state with specific projection \(M_J\) of $J$ {(note that $\ket{\Omega _{M_J}}$ is not associated with any physical system-probe interaction)}, the resulting \(\mathcal{L}(\sigma)\) receives contributions only from bound or continuum states satisfying \(J \ge |M_J|\).
Based on this scheme, we employ a set of single-configuration source states \(\{\ket{ \Omega _{M_J}}\}\) with cascading \(M_J\) values to resolve the structural information of the bound states.
The inputs, outputs, and methods of this protocol are summarized in Algorithm~1.

\begin{algorithm}[H]
	\caption{Prescan for spectral information}
	
	{\rm {\bf Input:}}
	
	\ \ \ 1. Single-configurations $\{ \ket{\Omega _{M_J} } \}$ with cascading $M_J$ 
	
	\ \ \ 2. $\sigma _I$ fixed to be a small positive value
	
	\ \ \ 3. $\{ \sigma _R \}$ {\it below} the threshold
	
	{\rm {\bf Methods:}}
	
	\ \ \ 1. Hybrid computations of $\mathcal{L}(\sigma) $ with $x=0$
	
	\ \ \ 2. $M_J$ scanning
	
	\ \ \ 3. Peak-structure fitting for $\mathcal{L}_D(\sigma)$
	
	{\rm {\bf Output:}}
	
	\ \ \ Bound-state information $\{ E_n \}$ and $\{ J_n \}$
	
\end{algorithm}

In practice, we set \(x = 0\) and choose a small \(\sigma_I\) to ensure well-separated peaks in \(\mathcal{L}_D(\sigma)\), while sampling $\sigma _R$ below the threshold energy \(E_{\rm th}\).
We compute the LI [Eq.~\eqref{eq:result_1_straight}] as a function of \(\sigma_R\) for source states $\{ \ket{ \Omega _{M_J}} \}$ with \(M_J > 1\); for \(\sigma_R < E_{\rm th}\), the response \(\mathcal{L}(\sigma)\) receives contributions solely from the discrete component \(\mathcal{L}_D(\sigma)\) [Eq.~\eqref{eq:response_function_formula_1}]. Bound-state energies \(E_n\) are extracted from the locations of the resolved peaks. Repeating the calculation with the source states $\{\ket{ \Omega _{M_J-1}}\}$ reveals newly appearing states with \(J = M_J - 1\), whose eigenenergies are extracted analogously. Iterating this procedure within the cascading \(M_J\) scheme yields the energies $\{E_n\}$ and angular momenta $\{J_n\}$ of the bound states.

In {\it the second protocol}, we compute the coefficients \(\{R_n\}\), which quantify the contributions of discretized states to the response function induced by a physical probe (Algorithm~2). We employ the same parameter settings as in the first protocol, namely \(\{\sigma_R\}\), \(\sigma_I\), and \(x = 0\), except that we adopt the physical source state $\ket{\Omega} = V \ket{\Psi_0}$. The LI is evaluated within the hybrid framework [Eq.~\eqref{eq:result_1_straight}]. We note that, for \(\sigma_R < E_{\rm th}\), only the discrete contribution \(\mathcal{L}_D(\sigma)\) enters. With small \(\sigma_I\), the peaks of \(\mathcal{L}_D(\sigma)\) are sufficiently separated, allowing \(\{R_n\}\) to be extracted by fitting the Lorentzian line shape [Eq.~\eqref{eq:response_function_formula_1}]. Using the bound-state energies $\{E_n\}$ obtained from the first protocol, each peak fit reduces to a single-parameter determination of \(R_n\).

We note that \(R_n\) depends on the bound state \(\ket{\Psi_n}\) and the physical source state \(\ket{\Omega}\), i.e., $R_n = \langle \Omega | \Psi_n \rangle \langle \Psi_n | \Omega \rangle$.
This identity provides a direct cross-check of the hybrid framework for model systems of limited size. For large many-body systems, however, the direct classical evaluation of \(R_n\) is computationally demanding; by exploiting the complementary strengths of quantum and classical computers, the present hybrid framework offers a viable and scalable solution.

\begin{algorithm}[H]
	\caption{Solving $\{  R_n \}$}

	{\rm {\bf Input:}}
	
	\ \ \ 1. $ \ket{\Omega } = V \ket{\Psi _0}  $ 
	
	\ \ \ 2. $ \sigma _I $ fixed to be a small positive value
	
	\ \ \ 3. $\{ \sigma _R \}$ {\it below} the threshold
	
	\ \ \ 4. Bound-state energies $\{ E_n \}$
	
	{\rm {\bf Methods:}}
	
	\ \ \ 1. Hybrid computations of $\mathcal{L}(\sigma)$ with $x=0$
	
	\ \ \ 2. Peak-structure fitting for $\mathcal{L}_D(\sigma)$
	
	{\rm {\bf Output:}}
	
	\ \ \ Bound-state response $\{ R_n \}$ 
	
\end{algorithm}

In {\it the third protocol}, we compute the response function \(R(e)\) as a function of the excitation energy \(e\), using the physical source state \(\ket{\Omega }=V \ket{\Psi_0} \) together with the inputs \(\{E_n\}\) (equivalently \(\{e_n\}\)) and \(\{R_n\}\) obtained from the first two protocols. We summarize this protocol in Algorithm~3. In practice, we set \(x = E_0\) and choose a moderate \(\sigma_I\) (e.g., of the order of $e_{\rm th}$) when evaluating the LI over \(\sigma_R \in [e_{\rm th},\, e_{\rm max}]\), where \(e_{\rm max}\) is some cutoff energy. For \(\sigma_R \ge e_{\rm th}\), the LI receives contributions from both bound and continuum states [Eq.~\eqref{eq:response_function_formula_1}]. The continuum contribution \(\mathcal{L}_C(\sigma)\) is obtained by subtracting the discrete contribution \(\mathcal{L}_D(\sigma)\) from the LI, and \(R(e)\) is then extracted from \(\mathcal{L}_C(\sigma)\) using the integral inversion technique in Sec.~3.5 of Ref.~\cite{Efros_2007}.

\begin{algorithm}[H]
	\caption{Solving $ R(e)$}
	
	{\rm {\bf Input:}}
	
	\ \ \ 1. $ \ket{\Omega } = V \ket{\Psi _0}  $ 
	
	\ \ \ 2. $ \sigma _I $ fixed to be a moderate positive value
	
	\ \ \ 3. $\{ \sigma _R \}$ values {\it above} threshold
	
	\ \ \ 4. Bound-state energies $\{ E_n \}$
	
	{\rm {\bf Methods:}}
	
	\ \ \ 1. Hybrid computations of $\mathcal{L} (\sigma)$ with $x=E_0$
	
	\ \ \ 2. Separating the $\mathcal{L}_C(\sigma) $ from $\mathcal{L}(\sigma)$
	
	\ \ \ 3. Integral inversion for $\mathcal{L}_C(\sigma)$ 
	
	{\rm {\bf Output:}}
	
	\ \ \ Continuous-state response $ R(e) $ 
	
\end{algorithm}

%{\color{gray} \bf Comment of the method; analysis of the gate and qubit cost.--}

%{\color{gray} Gate and qubit cost}, {\color{gray} Computational resources}, {\color{gray} Gaussian Integral transformation}

\begin{figure*}[ht]
	\centering
	\includegraphics[width=0.97\linewidth]{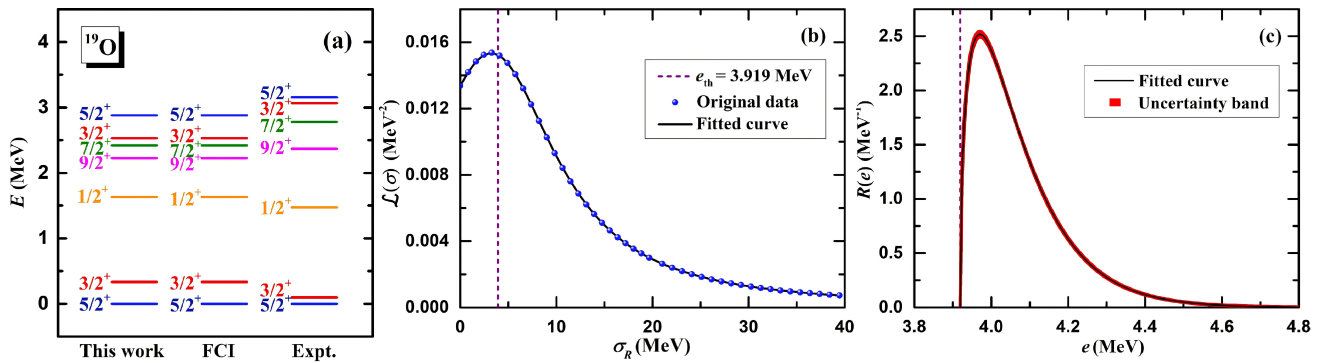}
	\caption{(a) Excitation spectrum of \({}^{19}\mathrm{O}\). The total angular momentum and parity are shown with each state. The results from the FCI calculations on classical computers and from the experiment \cite{NNDC2022} are also shown for comparison. (b) The LI as a function of $\sigma _R$ of \({}^{19}\mathrm{O}\). 
		(c) Response function $R(e)$ of \({}^{19}\mathrm{O}\) as a function of the excitation energy $e$. See text for more details.
	}
	\label{fig:total_results}
\end{figure*}

{\bf Results and discussions.--}
We demonstrate the framework by computing the bound-state structure and response function of \({}^{19}\mathrm{O}\) using a nonrelativistic two-body Hamiltonian formulated in a three-dimensional harmonic-oscillator basis~\cite{suhonen2007nucleons}. 
We retain the SP basis in the $0d_{5/2}1s_{1/2}$ {orbitals} and exclude those in the $0d_{3/2}$.
The few-body matrix elements of the Hamiltonian are available in Table~S3 of Ref.~\cite{Du:2025lhq}; these elements are derived from realistic nuclear interactions~\cite{Shin:2024zpe} via a transformation from the coupled basis to the SP basis~\cite{suhonen2007nucleons}. 
The Hamiltonian is block encoded into a quantum circuit employing the input scheme of Ref.~\cite{Du:2025lhq}.
Although the present calculations are performed in a limited model space, the same approach can be extended to no-core shell model calculations~\cite{Navratil:2000ww,Navratil:2000gs,Barrett:2013nh} and to BLFQ studies of structure and dynamics in QCD systems \cite{Vary:2025yqo,Honkanen:2010rc,Vary:2009gt,Wiecki:2014ola,Li:2015zda,Jia:2018ary,Du:2019ips,Lan:2019vui,Qian:2020utg}.

%{\color{magenta} \it Hybrid calculations are performed.-}
We evaluate the CMs employing the IBM Qiskit \cite{Qiskit} statevector simulator in noiseless mode and validate them against independent classical calculations. On future fault-tolerant quantum hardware, these moments can be obtained using standard quantum algorithms, such as the Hadamard test~\cite{nielsen2010quantum}. The LI is subsequently evaluated on a classical computer from the CMs according to Eq.~\eqref{eq:result_1_straight}.

%{\color{magenta} \it Spectral results are obtained and presented.-}
We apply the first protocol to achieve the bound-state spectrum of \({}^{19}\mathrm{O}\) applying a set of many-body states. 
As stated in Algorithm 1, these states are of cascading $M_J$ values, each of which is of a single configuration that can be prepared on the quantum computer in a straightforward manner.
The resulting excitation spectrum is shown in Fig. \ref{fig:total_results}a, together with their $J$ values and parities. These results agree with those obtained from classical calculations based on the full-configuration interaction (FCI) approach \cite{hjorth2017advanced}. We expect better agreement with the experiment \cite{NNDC2022} when we further include the SP bases in the $0d_{3/2}$ {shell}.

Next, we demonstrate the protocols for solving the LI and the response function. To this end, we elect a simple source state $\ket{\Omega}$ that is chosen to be of an equal-weight combination of the configurations $(1s_{\frac{1}{2},-\frac{1}{2}})  (1s_{\frac{1}{2},\frac{1}{2}})   (0d_{\frac{5}{2},\frac{1}{2}}) $, $(1s_{\frac{1}{2},\frac{1}{2}}) (0d_{\frac{5}{2},-\frac{5}{2}}) (0d_{\frac{5}{2}, \frac{5}{2}})$, and $ (0d_{\frac{5}{2},-\frac{5}{2}})  (0d_{\frac{5}{2},\frac{1}{2}})  (0d_{\frac{5}{2},\frac{5}{2}} ) $. This source state is obtained from the action of a test system-probe interaction $V$ (taken to be unitless) on the ground state $\ket{\Psi _0}$ of ${}^{19}\mathrm{O}$. 
We implement the oracle $W_{\Omega}$ to encode $\ket{\Omega}$, where, in this work, the circuit of $W_{\Omega}$ is realized by the protocol of initial state preparation in Qiskit. 

Based on $\ket{\Omega}$ and the spectral information, we apply Algorithm 2 to obtain $\{R_n\}$.
Then, we apply Algorithm 3 to obtain $\mathcal{L}_C(\sigma )$, based on which we obtain the response function $R(e)$ for $e \geq e_{\rm th}$. Here, we compute the threshold energy for the neutron separation to be $e_{\rm th} = 3.919$ MeV from the ground state energies of ${{}^{18}}$O and ${}^{19}\mathrm{O}$ obtained from our hybrid framework, which are $E_0({}^{18}\mathrm{O})= -11.155 $ MeV and $E_0({}^{19}\mathrm{O}) =-15.074$ MeV, respectively. This separation energy compare well with the experiment \cite{NNDC2022}, which is 3.956 MeV.

We present the LI as a function of $\sigma _R \geq 0$ in Fig. \ref{fig:total_results}b, where $\sigma _I  $ is fixed to be 8 MeV for illustration. According to Eq. \eqref{eq:response_function_formula_1}, only the bound states contribute to the LI below the threshold, while both the bound and continuum states contribute above it.  The $R(e)$ results for $e\geq e_{\rm th}$ are presented in Fig. \ref{fig:total_results}c. We find that the $R(e)$ results converge well. We tested that the converged results are independent of $\sigma _I$ over the range $\sigma _I \in [{5, 14}]$ MeV {based on the $\mathcal{L}_C(\sigma)$ results with $\sigma_R \in [e_{\rm th}, 40]$ MeV. The associated fitting uncertainties are indicated by the uncertainty band.} Furthermore, as a validity check, we reconstruct the LI according to Eq. \eqref{eq:response_function_formula_1} based on the fitted results of $R(e)$, $\{R_n\}$, and $\{E_n\}$ in Fig. \ref{fig:total_results}b. The reconstructed LI agrees with the input data.

{\bf Conclusions and outlook.--}
We introduce a quantum-classical framework for computing response functions and the full bound-state spectra of general many-fermion systems. 
This framework is based on the LIT and a new input scheme that enables efficient and practical circuit construction for block-encoding second-quantized Hamiltonians. 
We propose a hybrid scheme to compute the LI using a limited set of CMs evaluated on quantum computers. These CMs are post-processed on classical computers to reconstruct the LI. 
Based on the LI, we further introduce three protocols for extracting bound-state structural information and response functions of many-fermion systems. 
Our framework is practical and scalable, opening a new avenue for addressing various many-fermion structure and dynamics problems in quantum chemistry, nuclear physics, and field theories on future fault-tolerant quantum computers.

As a demonstration, we apply the framework to \({}^{19}\mathrm{O}\) using a realistic internucleon interaction. We obtain the eigenenergies and total angular momenta of the full bound-state spectrum, which show good agreement with the FCI results from classical calculations. We also compute the response function for a test probe in this work. With realistic system-probe interactions, such response functions provide reaction cross sections when combined with the corresponding kinematics.

Going forward, we plan to develop an efficient scheme for constructing the source-state oracle \(W_\Omega \), including systematic circuit designs for a wide class of system-probe interaction operators, together with efficient algorithms for ground-state preparation \cite{Lin_2020,PRXQuantum.3.040305}. Building on the hybrid framework introduced here, and incorporating the Hamiltonian input scheme for many-boson systems \cite{Du:2024ixj}, these developments will complete the toolkit for computing response functions of general many-body systems across various fields. Furthermore, we aim to simplify circuit designs (e.g., via quantum machine learning \cite{Biamonte_2017,Wang_2024}), thereby enabling demonstrations of the framework on NISQ hardware~\cite{Preskill2018quantumcomputingin} for simplified model problems.

{\bf Acknowledgments.--}
We are grateful for fruitful discussions with Xilin Zhang. JPV and CY are thankful for discussions with Pieter Maris and Peter Love. WD acknowledges discussions with Andrey M. Shirokov and Peng Yin. This work was supported in part by the U.S. DOE Grant DE-SC0023707 under the Office of Nuclear Physics Quantum Horizons program for the ``Nuclei and Hadrons with Quantum Computers (NuHaQ)” project. This project was also supported in part by NSF Grant No. 2435255 (NQVL-QSTD: Q-BLUE). WD is supported by startup funding from the Chinese Academy of Sciences.

%\newpage
%\normalem
\bibliography{apssamp}

\end{document}